\documentclass[runningheads]{llncs}
\usepackage{graphicx}

\usepackage{xcolor}
\usepackage[noend,linesnumbered]{algorithm2e}
\usepackage{amsfonts}
\usepackage{subcaption}
\usepackage{tikz}
\usepackage{pgfplots}
\usetikzlibrary{patterns}

\newcommand{\figref}[1]{Fig.~\ref{Fi:#1}}

\newcommand{\eqref}[1]{Eqn.~(\ref{Eq:#1})}

\newcommand{\sectref}[1]{Section~\ref{Se:#1}}

\newcommand{\algref}[1]{Alg.~\ref{Alg:#1}}
\newcommand{\algline}[1]{(Line~\ref{Line:#1})}
\newcommand{\alglines}[2]{(Lines~\ref{Line:#1}--\ref{Line:#2})}

\newcommand{\Omit}[1]{}
\newcommand{\True}{\ensuremath{\mathit{true}}\xspace}
\newcommand{\False}{\ensuremath{\mathit{false}}\xspace}
\newcommand{\clause}{\ensuremath{\omega}}
\newcommand{\Nat}{\ensuremath{\mathbb{N}^{+}}}
\newcommand{\EM}[1]{{\em #1}}
\newcommand{\toolname}{{\sf Open-WBO-Inc}\xspace}
\newcommand{\maxroster}{{\sf maxroster}\xspace}
\newcommand{\qmaxsat}{{\sf QMaxSAT}\xspace}
\newcommand{\satj}{{\sf SAT4J}\xspace}

\newcommand{\wpm}{{\sf WPM3}\xspace}
\newcommand{\maxhs}{{\sf MaxHS}\xspace}
\newcommand{\gtelsu}{{\sf apx-weight}\xspace}
\newcommand{\bmo}{{\sf apx-subprob}\xspace}

\newcommand{\pp}[1]{\vspace*{2mm}\noindent {\bf #1.}}
\colorlet{sjcolor}{blue}
\colorlet{rmcolor}{green}

\begin{document}
%
\title{Approximation Strategies for\\ Incomplete MaxSAT\thanks{Author names are in alphabetical order.}}
%
%
\author{Saurabh Joshi\inst{1} \and Prateek Kumar\inst{1} \and Ruben Martins\inst{2} \and  Sukrut Rao\inst{1}}
\authorrunning{S. Joshi, P. Kumar, R. Martins and S. Rao}
%
%
\institute{Indian Institute of Technology Hyderabad, India \\
\email{\{sbjoshi,cs15btech11031,cs15btech11036\}@iith.ac.in} \\
\and Carnegie Mellon University, USA \\
\email{rubenm@andrew.cmu.edu}}

\maketitle              
\begin{abstract}
Incomplete MaxSAT solving aims to quickly find a solution that attempts to minimize the sum of the weights of the unsatisfied soft clauses without providing any optimality guarantees. 
	In this paper, we propose two approximation strategies for improving incomplete MaxSAT solving. In one of the strategies, we cluster the weights and approximate them with a representative weight. In another strategy, we break up the problem of minimizing the sum of weights of unsatisfiable clauses into multiple minimization subproblems. Experimental results show that approximation strategies can be used to find better solutions than the best incomplete solvers in the MaxSAT Evaluation 2017.


\end{abstract}
\section{Introduction}

Given a set of Boolean constraints in a conjunctive normal form (CNF), the problem of Maximum Satisfiability (MaxSAT) asks to provide valuation of variables
so that maximum number of constraints are satisfied. These constraints can be assigned weights to prioritize some set of constraints over others, which would
give rise to a weighted MaxSAT problem where the goal is to find a valuation which maximizes the sum of the weights of the satisfied constraints. 
Any improvements in MaxSAT solving have a huge impact because many real world problems can
be encoded as MaxSAT problems (e.g.,~\cite{upgrade,bugassist,malware}).

Often, the application may be able to tolerate a suboptimal solution but requires
this solution to be computed in a very short amount of time. For such cases, it
tremendously helps if there are techniques and tools that can very quickly find a solution which is
close enough to an optimal solution. Incomplete MaxSAT solvers~\cite{maxroster,wpm3,dist,ccls,ccehc,deci} strive
to find a good solution in a limited time frame. The solution, thus provided, need not be
an optimal one. Therefore, for improvement, we need to develop tools and techniques that 
can find better solutions (closer to an optimal solution) in the same time frame.

As part of this paper, we contribute the following:
\begin{itemize}
	\item An approximation strategy based on weight relaxation (\sectref{weightrelax}), which modifies the weights of the clauses in a manner so that it is
		easier for the solver to find a solution quickly.
	\item An approximation strategy which breaks up the problem of minimizing the
		sum of weights of unsatisfied clauses into multiple minimization
		subproblems and attempts to minimize these subproblems in a greedy order
		(\sectref{weightrelax}). This strategy can also be combined with the weight
		relaxation strategy.
	\item Empirical results on how the accuracy of the solver gets affected
		as we vary the weight relaxation parameter (\sectref{results}).
	\item An implementation of these strategies using the {\sf Open-WBO} framework. We also
		demonstrate the advantage of these approximation strategies by showing its
		prowess against state-of-the-art incomplete MaxSAT solvers (\sectref{results}).
\end{itemize}


\section{Preliminaries}

Let $x$ be a Boolean variable which can take values \True or \False. A literal $l$ is a variable $x$ or its negation $\neg x$. 
A clause $\clause$ is a disjunction of literals and a formula $\varphi$ is a conjunction of
clauses. 
Notationally, we will treat a clause
$\clause$ and a formula $\varphi$ as sets containing literals and clauses respectively. 

An assignment $\nu$ maps variables to either \True or \False. An assignment is
said to satisfy a positive literal $x$ (resp. a negative literal $\neg x$) if
$\nu(x)=\True$ (resp. $\nu(x)=\False$). 
A clause is said to be satisfied if at least
one of its literals is satisfied. A formula is said to be satisfied by an assignment if
all of its clauses are satisfied by the assignment.
A formula is called \EM{satisfiable} if there exists a satisfying assignment for that formula, otherwise it is called \EM{unsatisfiable}.
Boolean satisfiability problem (SAT) asks to find a satisfying assignment (i.e., model) to a formula. Maximum satisfiability (MaxSAT) problem is an optimization version where the goal is to find an assignment which satisfies the maximum number of clauses of a formula. 
In a partial MaxSAT problem, a partition of $\varphi$ is given as two mutually exclusive
sets $\varphi_h$ (\EM{hard} clauses) and $\varphi_s$ (\EM{soft} clauses), where the goal is to satisfy all the clauses in $\varphi_h$~\footnote{For simplicity, we will assume that $\varphi_h$ is always satisfiable.} while maximizing the number of clauses satisfied in $\varphi_s$. 
Let $weight:\;Clauses \rightarrow \Nat $ be a map from a set of clauses to positive integers.
In a partial weighted MaxSAT problem, the goal is to find an assignment that maximizes the sum of weights of the satisfied soft clauses. 
From now on, we will refer to a weighted partial MaxSAT problem as MaxSAT.

A clause $\clause$ can be relaxed by adding a relaxation variable $r$ so that the relaxed clause becomes  $\clause \cup \{r\}$. The relaxed clause can be satisfied by either
satisfying the original clause or its relaxation variable. For a formula $\varphi$, when
all of its soft clauses are relaxed, we will denote it as $\varphi^r$. We define the cost of a
relaxation variable $r$  to be the weight of the clause that it relaxed,
$cost(r)=weight(\clause)$. The cost of an assignment $\nu$ is defined as $cost(\nu)=\sum _{r_i:\nu(r_i)=1} {cost(r_i)}$. The goal of MaxSAT is to find a satisfying assignment with the minimum cost.

\section{Approximation Strategies}

In this section, we describe two approximation strategies that can allow MaxSAT 
algorithms to converge faster to lower cost solutions. Note that the best 
model found by approximation strategies is not guaranteed to be an optimal 
solution of the original MaxSAT formula.

\pp{Weight-based approximation\label{Se:weightrelax}}
Let $P_m(\varphi_s)=\{c_1,\dots,c_m\}$ be a partition of $\varphi_s$ into $m$ mutually exclusive sets $c_1,\dots,c_m$ such that $\bigcup _{ 1 \leq i \leq m} c_i = \varphi_s$
and $\forall _{i\neq j} : c_i \cap c_j = \emptyset$. We will call sets $c_1,\dots,c_m$ as clusters of the partition.

\DontPrintSemicolon
\begin{algorithm}[t]
	\caption{Partitioning and weight approximation}
	\label{Alg:partitioning}
	\begin{scriptsize}
	\SetKwInOut{Input}{Input}\SetKwInOut{Output}{Output}
	\Input{Formula $\varphi_s$, Map $weight$, partitioning parameter $m$}
	\Output{Partition $P(m)$, new weight map $weight_m$}
	$n \gets |\varphi_s|$\;
	sort clauses of $\varphi_s$ in the ascending order of weights\;\label{Line:sort}
	\For{$i \gets 1$ \KwTo $n-1$} 
	{
		$diff_i \gets weight(\clause_{i+1})-weight(\clause_i)$\label{Line:diff}\;
		}

		$\langle i_1,\dots,i_{m-1} \rangle \gets$  sorted indices where top $(m-1)$ difference  $diff_i$ occurs\label{Line:topindices}\;
	
			$c_1 \gets \{\clause_{1},\dots,\clause_{i_1}\}$\label{Line:clusterstart}\;
		\For{$j \gets 2$ \KwTo $m-1$}
		{
			$c_j \gets \{\clause_{i_{j-1}+1},\dots,\clause_{i_j}\}$\;
			}
			$c_m \gets \{\clause_{i_{m-1}+1},\dots,\clause_{n}\}$\; \label{Line:clusterend}
			$P(m) \gets \{c_1,\dots,c_m\}$\;
	\ForEach{$c_i \in P(m)$ \label{Line:representweight}}{
	    \ForEach{$\clause_j \in c_i$}
              {
		      $weight_m[\clause_j]\gets RepresentativeWeight(c_i)$\; \label{Line:mean}
              }
}
	\Return $\langle P(m), weight_m \rangle$\;
\end{scriptsize}
\end{algorithm}

\DontPrintSemicolon
\SetKwFunction{SAT}{SAT}
\SetKwFunction{encodeCNF}{CNF}
\SetKwData{result}{model to}
\SetKwData{sat}{SAT}
\SetKwData{unsat}{UNSAT}
\SetKwData{true}{true}
\SetKwData{st}{status}
\SetKwData{model}{model}
\SetVlineSkip{1pt}
\begin{algorithm}[!t]
	\begin{scriptsize}
  \caption{Linear search Sat-Unsat algorithm for MaxSAT}\label{Alg:linearms}
  \KwIn{Formula $\varphi^r$, weight maps $weight_m$, and $weight$}
  \KwOut{model to $\varphi$}
  $(\model, \mu, \varphi_W) \gets (\emptyset, +\infty, \varphi^r)$\;
  $\st = \sat$\;
  \While{$\st = \sat$}{
   $(\st, \nu) \gets \SAT(\varphi_W)$\;
   \If{$\st =$ \sat}{
	   \If{$cost(\nu) < cost(\model)$ \label{Line:islessercost}}{
		   $\model \gets \nu$\label{Line:updatemodel}\;
   }
   $\mu \gets cost_m(\nu)$\label{li:linearms:ub}\;
   $\varphi_W \gets \varphi_W \cup \{\encodeCNF( (\sum_{r\in V_R} (cost_m(r) \cdot r)) \leq \mu-1 )\}$\label{Line:targetcost}\;
   }
  }
  \Return{\model}
\end{scriptsize}
\end{algorithm}

Given a formula $\varphi_s$, \algref{partitioning} partitions the clauses into clusters as follows.
All soft clauses are sorted by their weights \algline{sort}. Then, differences in weights between two consecutive
clauses are calculated \algline{diff}. $m-1$ indices are picked where the weight differences are amongst the top $m-1$ weight
differences \algline{topindices}. These indices are used as boundaries to create clusters \alglines{clusterstart}{clusterend}.
This way of clustering is similar to single-link agglomerative clustering~\cite{divisiveclustering}.
Finally, a new weight map $weight_m$ is created, where all the clauses in the same cluster get the same weight \alglines{representweight}{mean}.
%
$RepresentativeWeight$ \algline{mean} indicates any representative weight for the cluster.
In this paper, we use the \EM{arithmetic mean} of the weights of the clauses in a cluster as the representative weight. In principle, other representative
weights can also be chosen which may have different effect on how much an algorithm can
deviate from finding the minimum cost assignment. It is redundant to have $m>\#weights$, where $\#weights$ are different number of weights, because for $m\geq \#weights$, $weight=weight_m$. \algref{partitioning} can be combined with any search algorithm and as $m$ increases the deviation of the search algorithm from an optimal solution decreases. If $m=0$ it is assumed that no partitioning is done.

There are encodings which perform better when $\#weights$ is small~\cite{gte,minisat+}. Such encodings can benefit from approximation of weights because it results in a smaller size formula when converted to CNF. 
This can be used with a cost minimization algorithm for MaxSAT such as the linear 
search Sat-Unsat algorithm~\cite{sat4j,qmaxsat} shown in \algref{linearms}. 
In this algorithm, all the clauses in $\varphi_s$ are initially relaxed, and the set of corresponding relaxation variables is denoted as $V_R$. 
A working formula $\varphi_W$ is initialized with the relaxed formula $\varphi^r$. 
The cost of an empty \textsf{model} is assumed to be $+\infty$.
 Our primary goal is to find a satisfying assignment $\nu$ to $\varphi$
with the minimum $cost(\nu)$. \algref{linearms} iteratively asks a SAT solver if there is
a satisfying assignment, with its cost at most $\mu-1$ \algline{targetcost}. 
The approximation comes from \algref{linearms} using $cost_m$ instead of $cost$ to encode a pseudo-Boolean (PB) constraint
that restricts the cost of relaxation variables being set to \True \algline{targetcost}. Since $cost_m$ is an
approximation of $cost$, minimizing $cost_m$ does not necessarily translate to minimization w.r.t. $cost$. Therefore,
we update \textsf{model}, only when a satisfying assignment indeed reduces the previous value of $cost(\mathsf{model})$ \alglines{islessercost}{updatemodel}. 

\DontPrintSemicolon
\SetKwFunction{SAT}{SAT}
\SetKwFunction{encodeCNF}{CNF}
\SetKwData{result}{model to}
\SetKwData{sat}{SAT}
\SetKwData{unsat}{UNSAT}
\SetKwData{true}{true}
\SetKwData{st}{status}
\SetKwData{model}{model}
\SetVlineSkip{1pt}
\begin{algorithm}[t]
\begin{scriptsize}
  \caption{Clustering-based algorithm for MaxSAT}\label{Alg:bmo}
  \KwIn{$\varphi = \varphi_h \cup \varphi_s$, weight maps $weight$ and $weight_m$,
  Partition $P(m)$}
  \KwOut{model to $\varphi$}
  $(\model, \vec{\mu}, \varphi_W, {\mathcal C}) \gets (\emptyset, \vec{+\infty}, \varphi^r, P_m(\varphi_s)))$\;
  \ForEach{$c_i \in {\mathcal C}$ in the descending order of
  $weight_m(c_i)$\label{Line:descendingweights}}{
    $V_i \gets V_R \cap c_i$\label{Line:clusterrelax}\;
   $\st = \sat$\;
  \While{$\st = \sat$}{
   $(\st, \nu) \gets \SAT(\varphi_W)$\;
   \eIf{$\st =$ \sat \label{Line:lsucluster}}
   {
   \If{$cost(\nu) < cost(\model)$\label{Line:iflesscost}}{
    $\model \gets \nu$\label{Line:updatecost}\;
   }
  $\mu_i \gets \vert \{r \in V_i ~\vert~ \nu(r) = 1\} \vert$\label{Line:cardcost}\;
   $\varphi_W \gets \varphi_W \cup \{\encodeCNF( \sum_{r\in V_i} r \leq \mu_i-1
  )\}$\label{Line:cardmin}\;
   }
  {\label{Line:clusterunsat}
  $\varphi_W \gets \varphi^r$\label{Line:reset}\;
  \ForEach{$c_j \in \mathcal C$ such that  $weight_m(c_j) \geq weight_m(c_i)$}{
  $\varphi_W \gets \varphi_W \cup \{\encodeCNF( \sum_{r\in V_j} r \leq \mu_j
  )\}\label{Line:cardinality}$
   }
  }
  }
  }
  \Return{\model}
\end{scriptsize}
\end{algorithm}

\pp{Approximation via subproblem minimization}
\algref{bmo} proceeds in a greedy manner by processing each cluster in the descending order of its representative weight \algline{descendingweights}. $V_i$ indicates a set of relaxation variables
corresponding to the clauses in $c_i$ \algline{clusterrelax}. 
Minimization of the cost of a satisfying assignment is divided in subproblems by minimizing the number of unsatisfied clauses in clusters, starting from highest representative weight to the lowest \algline{descendingweights}.
For each cluster, the number of unsatisfied clauses are minimized by iteratively reducing the upper bound $\mu_i$ on the number of relaxation variables in $V_i$ that can be set to \True \alglines{cardcost}{cardmin}.
In the process, the minimum cost assignment seen so far is recorded \alglines{iflesscost}{updatecost}.
Since within any cluster, all the clauses have the same $weight_m$, only
cardinality constraints are used to restrict the number of unsatisfied clauses within
$\mu_i$ \algline{cardmin}. Once $\mu_i$ can not be reduced further, it is frozen
by  adding upper bound $\mu_i$ for all the cluster seen so far \alglines{clusterunsat}{cardinality}.
Since the minimization is done locally as a minimization subproblem at a cluster level,  rather than looking at the
whole formula, this procedure is not guaranteed to converge to a globally optimum solution.

\section{Related Work\label{Se:related}}

Approaches for incomplete MaxSAT solving can primarily be divided into two categories:
(i) stochastic MaxSAT solvers~\cite{dist,ccls,ccehc,deci,ramp} and (ii) complete
Max\-SAT solvers that can find intermediate solutions~\cite{qmaxsat,sat4j,open-wbo,wpm3,maxhs,lmhs}.

\pp{Incomplete MaxSAT} Stochastic solvers start by finding a
random assignment $\nu$ for $\varphi$. Since this assignment is unlikely to
satisfy all clauses in $\varphi$, they choose a clause $\clause_i$ that is unsatisfied
by $\nu$ and flip the assignment of a variable in $\clause_i$ such that $\clause_i$ becomes
satisfied. When compared to local SAT solvers, stochastic MaxSAT solvers have
additional challenges since they must find an assignment $\nu$ that satisfies
$\varphi_h$ while attempting to minimize the cost of the unsatisfied soft clauses.
Stochastic MaxSAT solvers are particularly effective for random benchmarks but
their performance tends to deteriorate for industrial benchmarks. Since the
MaxSAT Evaluation 2017 (MSE2017)~\cite{mse17} did not contain any random instances, there
were no stochastic MaxSAT solvers in the MSE2017.

\pp{Complete MaxSAT} Complete solvers can often find intermediate
solutions to $\varphi$ before finding an optimal assignment $\nu$.
MaxSAT solvers based on linear search
algorithms~\cite{sat4j,qmaxsat,open-wbo} can find a sequence of intermediate solutions
that converge to an optimal solution. These solvers use PB constraints to
enforce convergence. While \satj~\cite{sat4j} uses specialized data
structures for PB constraints to avoid their conversion to CNF, other solvers
such as \qmaxsat~\cite{qmaxsat} convert the PB constraint into clauses using PB
encodings~\cite{adder,gte,minisat+}.
Some MaxSAT solvers which are based on the implicit hitting set approach~\cite{maxhs,lmhs}
maintain a lower and an upper bound on the values of the solution. These
solvers can also be used for incomplete MaxSAT since they are also able to find
intermediate solutions.
Another approach for complete MaxSAT solving is to use unsatisfiability-based
algorithms~\cite{oll,wpm3,maxino}. These algorithms use unsatisfiable subformulas 
to increase a lower bound on the cost of a solution until they find an optimal
solution. For weighted MaxSAT, these algorithms employ a stratified approach~\cite{stratified}
where they start by considering only a subset of the soft clauses with the largest
weights and iteratively add more soft clauses when the subformula becomes satisfiable.
An intermediate solution is found at each iteration. \wpm~\cite{wpm3}
is an example of an unsatisfiable-based solver that can be used for incomplete
MaxSAT and was the best incomplete MaxSAT solver in the MSE2016~\cite{mse16}.
%
\maxroster~\cite{maxroster} was the winner of the incomplete track for Weighted
MaxSAT in the MSE2017. It is a hybrid solver that combines an
initial short phase of a stochastic algorithm~\cite{ramp} with complete MaxSAT
algorithms~\cite{oll,qmaxsat}. 

\pp{Boolean Multilevel Optimization} The clustering-based algorithm
presented in~\algref{bmo}
is closely related to Boolean Multilevel Optimization (BMO)~\cite{bmo}. BMO
is a technique for identifying lexicographic optimization conditions, i.e. the
existence of an ordered sequence of objective functions. Let $M_i$ be the minimum weight
of soft clauses in a cluster $c_i$. Consider a sequence of clusters $c_1, \ldots,
c_m$ arranged in a descending order of $M_i$. A MaxSAT
formula is an instance of BMO if for every cluster $c_i$, $M_i$ is larger than
the sum of the weights of all soft clauses in clusters $c_{i+1}, \ldots, c_m$. If this condition
holds then the result of~\algref{bmo} is equivalent to solving a BMO formula.
However, when using the proposed clustering-based algorithm on partitions that do
not preserve the BMO condition, it is not guaranteed that the solution found
by~\algref{bmo} is an optimal solution for $\varphi$. Our approach
differs from previous complete approaches in using approximation strategies
that do not preserve optimality but are more likely to converge faster to a better
solution.


\section{Experimental Results\label{Se:results}}

To evaluate incomplete MaxSAT solvers we used the scoring mechanism from MaxSAT Evaluations 2017 (MSE2017)~\cite{mse17}. Given a formula $\varphi$, the 
score for a solver $\mathcal S$ is computed by the ratio of the cost (sum of weights of unsatisfied clauses) of the best 
solution known for $\varphi$, denoted as $best(\varphi)$,~\footnote{$best(\varphi)$ is the cost of the best solution found by any solver in this evaluation.} to the best cost found by
$\mathcal S$, denoted as $cost^{\mathcal S}(\varphi)$.~\footnote{We consider a score of 0
if $\mathcal S$ did not find any solution to $\varphi$.} The score for $\mathcal S$ 
for a set of $n$ benchmarks is given by the average score ($[0,1]$) as follows:

\begin{equation}
	{\sf score}(\mathcal S) = \frac{\sum_{i=1}^n\frac{best(\varphi_i)}{cost^{\mathcal S}(\varphi_i)}}{n}
\end{equation}

\noindent {\sf score}($\mathcal S$) shows how close on average is a solver $S$ to the best 
known solution. 


All the experiments were conducted on Intel\textsuperscript{\textregistered}
Xeon\textsuperscript{\textregistered} E5-2620 v4 processors with a memory limit of 32GB 
and time limits of $10$, $60$ and $300$ seconds. We have used a non-standard timeout of
$10$ seconds to show that approximation strategies can find good solutions very quickly. 
We used the $156$ benchmarks for incomplete 
MaxSAT from MSE2017~\cite{mse17}. Note that most of these benchmarks are challenging for complete solvers and have unknown optimal solutions. 

We have implemented all the algorithms presented in this paper in \toolname. \toolname is built on top of {\sf Open-WBO}~\cite{open-wbo} which uses Glucose~\cite{glucose} as the underlying SAT 
solver. We used Generalized Totalizer Encoding (GTE)~\cite{gte} and incremental Totalizer
encoding~\cite{incremental-tot} to translate PB constraints and cardinality constraints
into CNF, respectively.



We evaluated \toolname by conducting experiments that are designed to answer the 
following questions: (1) What is the impact of the number of clusters in the quality of the 
solution found by our approximation strategies? (2) How does \toolname compare against state-of-the-art incomplete MaxSAT solvers?

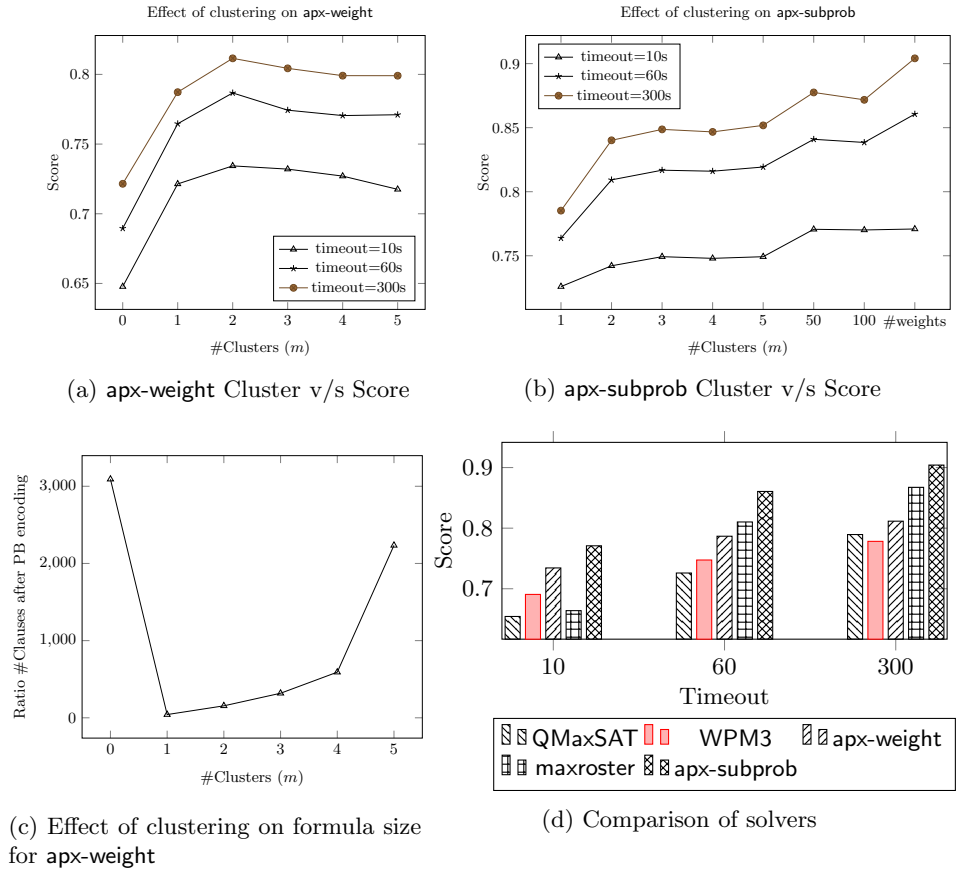
\begin{figure}[t]
	\begin{subfigure}[t]{.5\linewidth}
		\centering
		\begin{tikzpicture}[scale=0.64]
\begin{axis}[title=Effect of clustering on \gtelsu,
xlabel={\#Clusters ($m$)},
ylabel={Score},
ylabel style={yshift=-3.5mm},
legend entries={timeout=10s,timeout=60s,timeout=300s},
legend pos=south east,
]
\addplot [mark=triangle] table {plots/gte-lsu-10.dat};
\addplot  [mark=star] table {plots/gte-lsu-60.dat};
\addplot table {plots/gte-lsu-300.dat};
\end{axis}
\end{tikzpicture}
		\caption{\gtelsu Cluster v/s Score}
		\label{Fi:gtecluster}
	\end{subfigure}
	\begin{subfigure}[t]{.5\linewidth}
		\centering
		\begin{tikzpicture}[scale=0.64]
\begin{axis}[title=Effect of clustering on \bmo,
xlabel={\#Clusters ($m$)},
xtickmin=1,
xtickmax=\#weights,
xtick=data,
ylabel={Score},
x=1.05cm,
ylabel style={yshift=-3.5mm},
symbolic x coords={1,2,3,4,5,50,100,\#weights},
	xticklabels from table={plots/bmo-10.dat}{clusters},
legend entries={timeout=10s,timeout=60s,timeout=300s},
legend pos=north west,
]
	\addplot [mark=triangle] table {plots/bmo-10.dat};
\addplot [mark=star] table {plots/bmo-60.dat};
\addplot table {plots/bmo-300.dat};
\end{axis}
\end{tikzpicture}
		\vspace*{-4mm}
		\caption{\bmo Cluster v/s Score}
		\label{Fi:bmocluster}
	\end{subfigure}

\vspace*{3mm}
\begin{subfigure}[t]{.45\linewidth}
\centering
\begin{tikzpicture}[scale=0.66]
\begin{axis}[
xlabel={\#Clusters ($m$)},
ylabel={Ratio \#Clauses after PB encoding},
legend pos=south east
]
\addplot [mark=triangle] table {plots/gte-lsu-clauses.dat};
\end{axis}
\end{tikzpicture}
\vspace*{-3mm}
\caption{Effect of clustering on formula size for \gtelsu}
\label{Fi:clauses}
\end{subfigure}
\begin{subfigure}[t]{.55\linewidth}
	\centering
\begin{tikzpicture}
 \begin{axis}[
width=7.5cm,
height=4.2cm,
 ybar, 
enlargelimits=0.15, 
legend style={at={(0.5,-0.4)}, anchor=north, legend columns=3},
ylabel={Score}, 
xlabel={Timeout},
ylabel style={yshift=-4.5mm},
symbolic x coords={10,60,300}, 
xtick=data, 
bar width=.2cm,
 ] 
 \addplot [pattern=north west lines] table{plots/qmaxsat-time.dat};
 \addplot  table{plots/wpm3-time.dat};
 \addplot [pattern=north east lines] table{plots/gte-time.dat};
 \addplot [pattern=grid] table{plots/maxroster-time.dat};
 \addplot [pattern=crosshatch] table{plots/bmo-time.dat };
\legend{\qmaxsat,\wpm,\gtelsu,\maxroster,\bmo};
 \end{axis} 
\end{tikzpicture}
	\caption{Comparison of solvers}
	\label{Fi:comparision}
	\end{subfigure}
	\caption{Impact of clustering and comparison against state-of-the-art}
	\label{Fi:results}
\end{figure}

\pp{Impact of the number of clusters} We measure the impact of partitioning parameter $m$ on the accuracy
of the results.
\figref{gtecluster} shows the score of \algref{linearms} with GTE encoding (henceforth
called \gtelsu). \figref{gtecluster} shows that \gtelsu performs the worst when no partitioning is done. This is attributed to the fact that in the absence of 
any partitioning, the size of the underlying encoding is dictated by $\#weights$, where $\#weights$ are the number of different
weights in the weight map. \figref{clauses} shows a measure of increase in formula size as
$m$ varies. The $Y$-axis shows the ratio of the formula size after the PB encoding to the size
of the original input formula. Because of the weight-based approximation, the reduction on
the $\#weights$ leads to a smaller encoding, thus making it easier for the underlying SAT
solver. As $m$ increases, the possible deviation from
an optimal cost also decreases, thereby resulting in increased scores.
The degradation for larger $m$ is attributed to larger size of the formula. As the timeout is increased,
the score increases because \algref{linearms} has more time and can do more
iterations to reduce $cost_m(\mathsf {model})$.


\figref{bmocluster} shows that similar scoring trends are witnessed for \algref{bmo}
(henceforth called \bmo). As \bmo uses only cardinality constraints, the formula size is not much sensitive to $m$. 
As $m$ increases, the scores also improve, with the best scores
achieved when $m=\#weights$.
\bmo is guaranteed to find optimal solution only if BMO condition holds and $m=\#weights$.
Only 3 out of 156 benchmarks have the BMO condition and \bmo with $m = \#weights$ does not
terminate for any of them. However, \bmo using a 300 second time limit terminates for 94
out of 156 benchmarks which shows that \bmo quickly finds a good solution.
%



\pp{Comparison against state-of-the-art MaxSAT solvers}
We compared the best version of \toolname for weight-based approximation, \gtelsu with $m=2$, and subproblem minimization approximation, \bmo with $m=\#weights$, with \maxroster~\cite{maxroster}, \wpm~\cite{wpm3} and \qmaxsat~\cite{qmaxsat}. \maxroster and \wpm were 
the winners of the incomplete weighted category of the MSE2017 and MSE2016, 
respectively. \qmaxsat was placed second on the complete category of the MSE2017 and uses the algorithm described in \algref{linearms}.~\footnote{Even though \maxhs~\cite{maxhs} placed first in the complete weighted category of the MSE2017, its incomplete version is not as competitive as the other solvers~\cite{mse17}.
}

As shown in \figref{comparision}, for a $10$ seconds timeout, both \gtelsu and \bmo perform  better than all the other solvers with \bmo performing
the best. This demonstrates that approximation strategies are quite effective when we want
to quickly find a solution which is closer to an optimal solution. For $60$ and $300$ seconds timeout, \bmo performs the best with \maxroster being second and \gtelsu outperforming \wpm and \qmaxsat. 
Even though \gtelsu with $m=0$ performs worse than \qmaxsat, 
it outperforms \qmaxsat when clustering is used. \bmo outperforms all other solvers and these results prove the efficacy of approximation strategies with respect to the state-of-the-art in incomplete MaxSAT solving.

\section{Conclusion and Future Work\label{Se:conclusion}}
Approximation strategies, be it weight-based relaxation or subproblem minimization, are not guaranteed to find
an optimal solution even when unlimited time is given. However, they serve the purpose of quickly finding a
good solution. Our experiments have successfully demonstrated that with the right parameters, these 
strategies can outperform the best incomplete solvers. 
In future, we would like to explore the application of approximation strategies to
complete algorithms. In particular, progressively increasing the number of clusters and
using approximation strategies to find good initial upper bounds that can later be exploited by complete MaxSAT algorithms.


\section*{Acknowledgements}
This work is partially funded by ECR 2017 grant from SERB, DST, India, NSF award \#1762363 and CMU/AIR/0022/2017 grant. Authors would like to thank the anonymous reviewers for their helpful comments, and Saketha Nath for lending his servers for the experiments.

\bibliographystyle{plain}
\bibliography{biblio}
\end{document}